\documentstyle[12pt,epsf]{article}
\newcommand{\fcz}{$F_{C0}(q)$}
\newcommand{\fmo}{$F_{M1}(q)$}
\newcommand{\fct}{$F_{C2}(q)$}
\newcommand{\Aq}{$A(q)$}
\newcommand{\gen}{$G_{En}$}

\begin{document}
\newcommand{\fmsq}{$fm^{-2}$}
\newcommand{\gev}{$(GeV/c)^{2}$}
\newcommand{\mgcm}{$\frac{mg}{cm^{2}}$}
\newcommand{\qv}{$\vec{q}$}
\newcommand{\qsq}{$Q^2$}
\newcommand{\w}{$\omega$}
\newcommand{\om}{$\omega$}
\newcommand{\eps}{$\epsilon$}
\newcommand{\qve}{$q_{eff}$}
\newcommand{\dpp}{$\Delta p/p$}
\newcommand{\dth}{$\Delta \vartheta$}
\newcommand{\dph}{$\Delta \varphi$}
\newcommand{\ee}{$(e,e')$}
\newcommand{\eep}{$(e,e'p)$}   
\newcommand{\een}{$(e,e'n)$}   
\newcommand{\dee}{$D(e,e')$}
\newcommand{\deep}{$D(e,e'p)$}
\newcommand{\deen}{$D(e,e'n)$}
\newcommand{\heep}{$\vec{^3He}(\vec{e},e'p)$}
\newcommand{\heen}{$\vec{^3He}(\vec{e},e'n)$}
\newcommand{\deed}{$D(e,e'\vec d)$}
\newcommand{\epi}{$H(e,\pi^+)n e'$}
\newcommand{\epin}{$H(e,\pi^+ n)e'$}
\newcommand{\epn}{$D(e,pn)e'$}
\newcommand{\np}{$H(n,p)n$}
\newcommand{\pn}{$H(n,n)p$}
\newcommand{\hdpp}{$^{1}$H($\vec{\normalsize d}$,pp)n}          
\newcommand{\hdpn}{$^{1}$H($\vec{\normalsize d}$,pn)p}          
\newcommand{\hng} {$^{1}$H($\vec{n}$,$\gamma$)$^{2}$H}          
\newcommand{\dpg} {$^{1}$H($\vec{d}$,$\gamma$)$^{3}$He}         
\newcommand{\ddg} {$^{2}$H($\vec{d}$,$\gamma$)$^{4}$He}         
\newcommand{\dpn}{$^2$H($\vec{p},\vec{n})pp$}
\newcommand{\pvCss}{$^{14}C({\vec p},{\vec n})^{14}N(2.31)\:$}	
\newcommand{\pvCts}{$^{14}C({\vec p},{\vec n})^{14}N(3.95)\:$}	
\newcommand{\cpn}{$^{14}$C($\vec{p}$,n)$^{14}$N(2.31MeV)}
\newcommand{\fl}{$f_L$({\bf {q}},$\omega,E$)}
\newcommand{\ft}{$f_T$({\bf {q}},$\omega,E$)}
\newcommand{\rl}{$R_L(q,\omega)$}
\newcommand{\rt}{$R_T(q,\omega)$}
\newcommand{\slw}{$S_L^{\omega_{max}}(q)$}
\newcommand{\slq}{$S_L(q)$}
\newcommand{\slinf}{$\Sigma_L(q)$}
\newcommand{\stq}{$S_T(q)$}
\newcommand{\sigep}{$\sigma_{e-p}(q)$}
\newcommand{\eff}{$\eta(x,y,T_n)$}
\newcommand{\specf}{$S(\vec{k},E)$}
\newcommand{\xnm}{${\cal R}_{NM}$ }
\newcommand{\xa}{${\cal R}_{A}$ }
\newcommand{\xnmns}{${\cal R}_{NM}$}
\newcommand{\xans}{${\cal R}_{A}$}
\newcommand{\e}{$\epsilon_{1}$ \hspace{0.5mm}}
\newcommand{\ayy}{A$_{yy}$}      %
\newcommand{\ay}{A$_{y}$}        %
\newcommand{\ayd}{A$_{y}^d$}        %
\newcommand{\axx}{A$_{xx}$}      %
\newcommand{\azz}{A$_{zz}$}
\newcommand{\azx}{A$_{zx}$}
\newcommand{\Kyy}{$K_{y}^{y\prime}(0^{\circ})$ }	
\newcommand{\atl}{A$_{TL}$}    
\newcommand{\att}{A$_{TT}$}    
\newcommand{\gmn}{$G_{mn}$}
\newcommand{\gep}{$G_{ep}$}
\newcommand{\gmp}{$G_{mp}$}
\newcommand{\gc}{$G_{C}$}
\newcommand{\gm}{$G_{M}$}
\newcommand{\gq}{$G_{Q}$}
\newcommand{\fch}{$F_{c}^{H}$ }
\newcommand{\fmh}{$F_{m}^{H}$ }
\newcommand{\fche}{$F_{c}^{He}$}
\newcommand{\fmhe}{$F_{m}^{He}$}
\newcommand{\htt}{$^{3}$H}
\newcommand{\het}{$^{3}He$}
\newcommand{\he}{$^{4}He$}
\newcommand{\be}{$^{9}$Be}
\newcommand{\cc}{$^{12}C$}
\newcommand{\cf}{$^{14}C$}
\newcommand{\al}{$^{27}Al$}
\newcommand{\ca}{$^{40}Ca$}
\newcommand{\cae}{$^{48}$Ca}
\newcommand{\fe}{$^{56}Fe$}
\newcommand{\cu}{$^{65}$Cu}
\newcommand{\sn}{$^{120}$Sn}
\newcommand{\au}{$^{197}$Au}
\newcommand{\pp}{$^{208}$Pb}
\newcommand{\uu}{$^{238}$U}
\newcommand{\numa}{nuclear matter}
\newcommand{\nm}{nuclear matter}
\newcommand{\BaF}{BaF$_{2}\:$}
\newcommand{\effi}{$\eta (T_n,x,y)$}
\newcommand{\at}{$A^{-1/3}$}
\newcommand{\eone}{$\epsilon_1$}
\newcommand{\pone}{$^{1}P_{1}$}
\newcommand{\fsi}{final state interaction}
\newcommand{\mec}{meson exchange currents}
\newcommand{\dei}{$\Delta E_i$}
\newcommand{\de}{$\Delta E$}
\newcommand{\ef}{$E_{front}$}
\newcommand{\er}{$E_{rear}$}
\newcommand{\phm}{photomultiplier}
\newcommand{\ld}{$LD_{2}$}
\newcommand{\lh}{$LH_{2}$}
\newcommand{\tof}{time of flight}
\newcommand{\et}{{\em et al.}}
\newcommand{\bq}{\begin{eqnarray}}
\newcommand{\eq}{\end{eqnarray}}
\newcommand{\beq}{\begin{equation}}
\newcommand{\eeq}{\end{equation}}

\thispagestyle{empty}    

\begin{center}
\LARGE
Neutron charge form factor at large $q^2$ \\[1cm]
\large
 R. Schiavilla$^{ab}$ and I. Sick$^c$ \\[5mm]
\normalsize
$^a$Jefferson Lab, Newport News, Virginia, USA \\
$^b$Physics Department, Old Dominion University, Norfolk, Virginia, USA \\
$^c$Departement f\"ur Physik und Astronomie, Universit\"at Basel, Basel,  
Switzerland \\[2cm]
\begin{minipage}{12cm}
{
\small 
{\bf Abstract.} The neutron charge form factor $G_{En}(q)$ is determined
from an analysis of the deuteron quadrupole form factor \fct \ data.  Recent
calculations, based on a variety of different model interactions and
currents, indicate that the contributions associated with the uncertain
two-body operators of shorter range are
relatively small for \fct, even at large momentum
transfer $q$.  Hence, $G_{En}(q)$ can be extracted from \fct \
at large $q^2$ without undue systematic uncertainties from theory.
}
\end{minipage} 

\end{center}
\vspace{1cm}
{\bf Introduction.} 
Knowledge of the neutron charge form factor $G_{En}$ is of great 
importance for an understanding of its internal structure.  It is also
crucial for the calculation of nuclear charge form factors, since the
latter depend on both $G_{En}$ and the proton charge form
factor $G_{Ep}$.

Unfortunately, $G_{En}$ is still rather poorly known. The difficulties
encountered in measuring $G_{En}$ are twofold: since there are no free
neutron targets, $G_{En}$ has to be measured using composite systems,
and this leads to complications due to the presence of other nucleons.
In addition, the electron-neutron cross section is dominated by the
contribution from the magnetic form factor $G_{Mn}$, thus making a
determination of $G_{En}$ very difficult. 

The traditional approach to determine $G_{En}$ uses the deuteron 
structure function \Aq , to which the deuteron magnetic form factor,
and therefore $G_{Mn}$, contributes negligibly.  After removing,
via theoretical calculations, the effect of the deuteron structure
and the contributions to the scattering process from two-body currents,
subtraction of the $G_{Ep}$ contribution then allows one to extract
$G_{En}$.  This procedure is sensitive to systematic errors in the theory
used, particularly those associated with the modeling
of shorter-range two-body currents, which are still not very
well controlled.

As a consequence, the resulting values for $G_{En}$~\cite{Platchkov90,Galster71}
have fairly large uncertainties, and are limited to momentum transfers below 
$q^2$=16 fm$^{-2}$.  This poses serious problems for the calculation of 
form factors of light nuclei, which one would want to calculate for
the region covered by data, a region that extends to $q^2$=30--100 fm$^{-2}$.
To the extent that current parameterizations of nucleon form factors provide 
sensible
extrapolations for $G_{En}$ at large $q^2$, one must conclude that 
the contribution of $G_{En}$, which seems to fall off with increasing $q^2$ much 
more slowly than $G_{Ep}$, becomes very important at these large momentum 
transfers.

More recently, the exploitation of a new technique to determine $G_{En}$
has become practical: when performing an $(e,e^\prime n)$ coincidence
experiment using {\em polarized} electrons and when measuring the
{\em polarization} of the target nucleus or recoil neutron, it becomes
possible to measure an interference term $G_{En} G_{Mn}$.  This approach
removes the difficulty associated with $G_{Mn}$-dominance, and is much less
dependent on the nuclear structure of the target nucleus (deuteron or $^3$He).
Several experiments of this type have been performed
recently~\cite{Meyerhoff94a}--\nocite{Becker97,Becker99,Rohe99,Herberg99,Eden94,Ostrick99,Passchier99}\cite{Ahmidouch01}.
The resulting values for $G_{En}$ still have relatively large errors; 
they are, however, mainly statistical and thus can be reduced in the future 
using better technology.  The limit in $q^2$ is presently 17 fm$^{-2}$.
Two experiments are under way at JLAB to extend the
$q^2$-range~\cite{Day93x,Madey93x}. 
Within the error bars the available
results from the double-polarization experiments
agree  with the values determined from the deuteron
structure function $A(q)$.

{\bf Exploitation of the quadrupole form factor. } In this note, we again use
elastic electron-deuteron data to determine $G_{En}(q)$ at high momentum 
transfer.  The novel aspect of the present approach  consists
in exploiting the {\em quadrupole} form factor $F_{C2}(q)$ 
rather than the combination of monopole and quadrupole form 
factors represented by \Aq , as done in the past.  

When using elastic e-d scattering, two sources of theoretical uncertainty
must be considered, due to the model for the NN-interaction and the
contribution of two-body currents. We address the two-body currents first.  

\begin{figure}[htb]
\centerline{\mbox{\epsfysize=60mm\epsffile{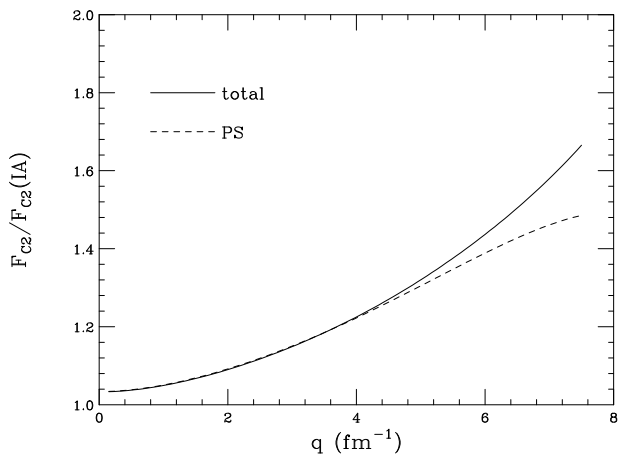}}}
\caption{Effect of the $\pi$-exchange two-body charge
operator (dashed) and that obtained by including
the remaining, shorter range, two-body contributions (solid).}
\label{mmecr}
\end{figure}

Calculations
of $F_{C2}(q)$ based on a variety of model interactions and currents
indicate that contributions from two-body currents are
relatively small, even at the high momentum transfers of
interest here.  This is consistent with the naive expectation
that, since $F_{C2}(q)$ involves an integral of the product
of deuteron S- and D-wave components with the spherical
Bessel function $j_2(qr/2)$, it is presumably less sensitive
to two-body currents, at least the short-range ones associated
with vector-meson exchanges and/or transition mechanisms such as, for
example, the $\rho \pi \gamma$ operator, whose contributions are
quantitatively rather uncertain.  This is illustrated in
Fig.~\ref{mmecr} where we show separately the contribution
associated with the $\pi$-exchange two-body charge operator,
as well as that including, in addition, the $\rho$-meson
and $\rho\pi\gamma$ charge operators.  The $\rho$-meson and
$\rho\pi\gamma$ contributions have opposite sign, and tend to cancel
each other.  As a result, the total two-body contribution
to \fct \ is dominated, up to $q^2 \simeq 40$ fm$^{-2}$, by the
long-range $\pi$-exchange operator.

In this context, it is worth noting that, while modern realistic 
interactions are essentially
phase-equivalent --- they all fit the Nijmegen data-base with a
$\chi^2$ per datum close to one --- they do differ
in the treatment of non-localities.  Some of them, like the
Argonne $v_{18}$ model~\cite{Wiringa95}, are local (in $LSJ$ channels),
while some others, like the CD-Bonn model~\cite{Machleidt96},
have strong non-localities.  In particular, the CD-Bonn interaction has a
non-local one-pion-exchange (OPE) component.  However, it has been known
for some time~\cite{Friar77}, and recently
re-emphasized by Forest~\cite{Forest00}, that the
local and non-local OPE interactions are related to each other
via a unitary transformation.  Therefore, the differences between
local and non-local OPE cannot be of any consequence for the prediction
of observables, such as the deuteron electromagnetic form
factors under consideration here, provided, of course, that
two-body currents generated by the unitary transformation are
also included.  This fact has been
demonstrated~\cite{Schiavilla01} in a calculation of
the deuteron structure function $A(q)$ and tensor observable $T_{20}(q)$,
based on the local Argonne $v_{18}$ and non-local CD-Bonn
models and associated (unitarily consistent) electromagnetic currents.
The remaining small differences between the calculated $A(q)$ and
$T_{20}(q)$ are due to the additional short-range non-localities
present in the CD-Bonn.  The upshot is that, provided
that consistent calculations --- in the sense above --- are performed,
present ``realistic'' interactions will lead to similar
predictions for deuteron electromagnetic observables, at
least to the extent that these are influenced predominantly
by the OPE component. This is especially true for the $F_{C2}$ form factor
for which the $\pi$-exchange contributions dominate.

Because of these considerations,
the theoretical uncertainties for
\fct \ are small (smaller than for \Aq), which allows us to 
to determine $G_{En}$ with smaller systematic errors and extend
our knowledge of it to larger $q$.  The use of \fct \ 
has now become possible with the measurements of the polarization observable 
$T_{20}(q)$ in electron-deuteron scattering.  With $T_{20}$ known up to 
$q^2$=40 fm$^{-2}$, the quadrupole form factor \fct \ can experimentally be
determined up to that $q$-value.

{\bf Experimental \fct .} In order to determine \fct , we have analyzed the 
{\em world} data on electron-deuteron elastic scattering
\cite{Abbott99} \nocite{Alexa99,Akimov79}
\nocite{Arnold75,Arnold87,Auffret85a,Benaksas66,Berard73a,Bosted90}
\nocite{Bumiller70,Buchanan65,Cramer85,Drickey62}
\nocite{Elias69,Friedman60,Galster71,Ganichot72,Goldemberg64}
\nocite{Honegger97,Grossetete66b}
\nocite{Martin77,Platchkov90,Rand67,Simon81,Stein66}--
\cite{Voitsekhovskii86}. Some 340 data points on e-d scattering are available 
for momentum transfers
below 65 fm$^{-2}$.
The cross sections and polarization observables are fitted with flexible
parameterizations for \fcz , \fmo \ and \fct \cite{Sick74}. The statistical 
errors of the data are calculated using the error matrix. The systematic errors,
which in general are the largest ones by far, have been
evaluated by changing each individual data set by the quoted error, and 
re-fitting the complete data set. 
The changes due to systematic errors  of the different, independent,
sets of data are evaluated separately, and added quadratically.
The resulting \fct \ is used below.

{\bf Determination of \gen .} In order to extract \gen \ we compare to the 
predictions for \fct \ from a
number of theoretical calculations.  These calculations all use NN potentials
that provide reasonably good fits to the modern
scattering data base, and consistent two-body currents.
We employ calculations using the 
Paris and Bonn-B potential by the Hannover group \cite{Buchmann96,Henning97}, 
the calculation of Forest and Schiavilla \cite{Forest01} based on the
Argonne V18 NN potential, and the results obtained recently by the Mainz
group \cite{Arenhoevel00} using the Bonn OBEPQ-B potential. 
We also employ the results of the calculation
by Van Orden, Gross and Devine \cite{VanOrden95} who use an OBE interaction
directly fit to the NN scattering data. 

While the first three calculations are based on an essentially non-relativistic 
framework (with relativistic corrections), the calculation  \cite{Arenhoevel00}
starts from a system of coupled
nucleon and meson fields and, by means of the Foldy-Wouthuysen transformation,
derives the non-relativistic limit including all the leading order
relativistic contributions.   The calculation of Van Orden {\em et al.}
starts from the Bethe-Salpeter equation, which has been reduced to a 
quasi-potential equation by assuming that one of the nucleons is on mass shell. 
This calculation is Lorentz covariant and gauge invariant. All calculations
include the relevant two-body terms.

In general, these calculations have used proton form factors as given by the
Hoehler parameterization \cite{Hoehler76}, which explains well the e-p scattering data up to 
the $q^2$ of interest here, including the recent $G_{Ep}/G_{Mp}$-data 
\cite{Jones00,Popischil00,Milbrath98,Sick01}. The calculations of refs. 
\cite{Arenhoevel00,VanOrden95}
have been carried out using the dipole form factor for the proton, which only
roughly reproduces the proton data; here we use the calculation of 
Arenhoevel \et \ \cite{Arenhoevel01} performed with the Hoehler form factors, 
while the calculation of van Orden \et\ 
has  been renormalized to the Hoehler proton form factor.  
 All calculations use the Galster \cite{Galster71} neutron charge form factor,
or the one by Hoehler, which is very close in the range of $q^2$ of interest. 

\begin{figure}[htb]
\centerline{\mbox{\epsfysize=70mm\epsffile{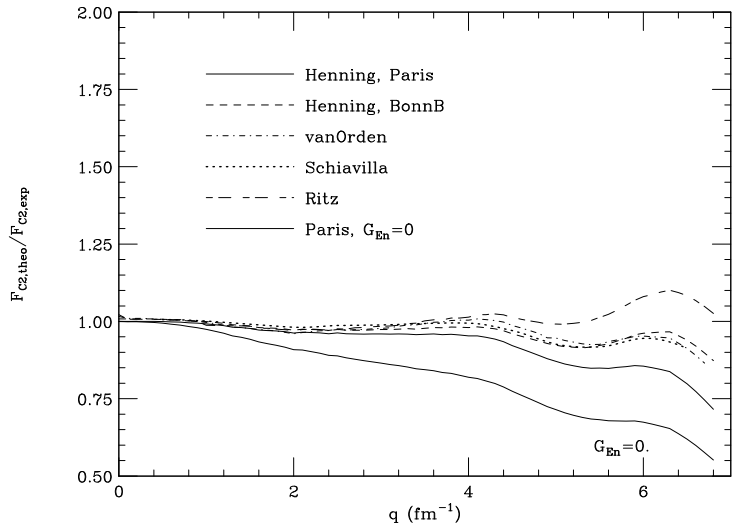}}}
\caption{The ratios of theoretical to experimental C2 form factors as a function
of momentum transfer.  For the Paris potential, we also give the ratio
obtained setting \gen =0.}
\label{rate2ms1}
\end{figure}

In fig. \ref{rate2ms1} we show the ratio of these theoretical \fct \ form 
factors to the experimental ones. 
This figure shows that for the C2 form factor the different theoretical predictions
are quite close.  The effect of \gen \ is appreciable at the higher momentum transfers,
large enough to be extracted despite the differences between the theoretical 
predictions. 

\begin{figure}[htb]
\centerline{\mbox{\epsfysize=90mm\epsffile{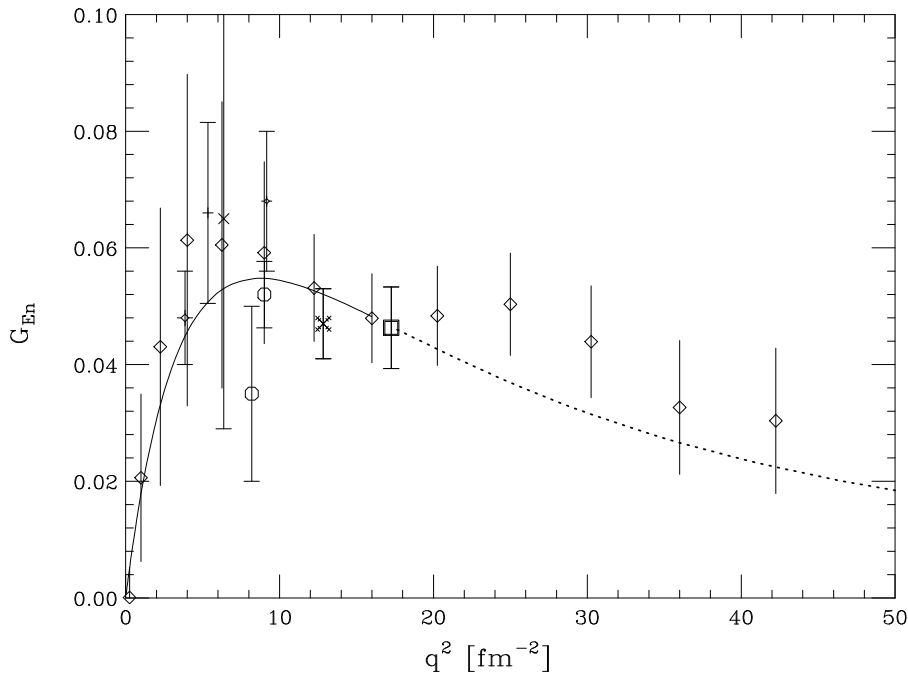}}}
\caption[]{The \gen \ extracted from the C2 data ($\diamond$). Also shown are the values 
obtained from double-polarization experiments, and the Galster parameterization 
with its extrapolation into the region not covered by previous 
experiments (dotted). }
\label{rate2ms2}
\end{figure}

In order to determine \gen , we use the following approach:
As the \lq\lq theoretical prediction\rq\rq we use the {\em average}
of the five calculations discussed 
above. For the \lq\lq theoretical  error bar\rq\rq we take the quadratically added 
deviation of the individual calculations from the average. The deviation of this
average from experiment we then take as an indication that the Galster (or Hoehler) 
\gen \ used in the calculation is not quite the correct one, and we determine
\gen \ to get perfect agreement between experiment and the theory average.
The resulting values of \gen , together with the error bars that include {\em both} the 
spread of the theoretical predictions {\em and} the experimental uncertainty on $F_{C2}$,
is shown in fig. \ref{rate2ms2}.

Figure \ref{rate2ms2} shows that the form factors extracted from the C2 deuteron
structure function are  reasonably
accurate in comparison with the results obtained from
double-polarization measurements, and they
agree with them  in the $q^2$-region 
of overlap. In comparison to the mean
values of \gen \ determined by Platchkov \et\ 
from the deuteron $A(q)$ structure function, the present results are somewhat
higher in the region above $q^2$=8 fm$^{-2}$, but compatible with them given the  
spread of the theoretical predictions available to Platchkov \et\ at the time.
The \gen \ extracted from the $C2$-data  have larger uncertainties at low $q^2$, 
where the $C0$ multipolarity dominates
the cross section and where the available $T_{20}$ data are not very accurate.  
There, the usage of $A(q)$ leads to  superior results.

The determination of \gen \ from $F_{C2}$ extends to larger
momentum transfer than all previous determinations, which were limited
to $q^2 \simeq 16$ fm$^{-2}$.  Somewhat surprisingly, the extrapolation
of the Galster parameterization beyond its limit of
validity ($q^2$=16 fm$^{-2}$) does quite well in reproducing the data. 
As pointed out above, double-polarization experiments presently under
way at JLAB are expected to provide data in this higher-$q^2$ region.  

{\bf Conclusions.} In this note, we have determined the neutron charge form 
factor \gen \ starting from the data on electron-deuteron elastic scattering.
Contrary to previous analyses, we use the deuteron 
{\em quadrupole} form factor, which is less sensitive to the 
short-range two-body currents that are not well under control.  We employ a 
representative selection of both non-relativistic and relativistic 
theoretical calculations to predict deuteron
structure functions and contributions of two-body currents, thus allowing 
use to produce a fair estimate of the theoretical uncertainties involved
in our procedure.  Using this approach, we for the first time provide 
data (other than upper limits \cite{Lung93}) for \gen \ at large $q^2$.
 
{\bf Acknowledgments.} 
RS is supported by the U.S. Department of Energy contract DE-AC05-84ER40150
under which the Southeastern Universities Research Association (SURA)
operates the Thomas Jefferson National Accelerator Facility, while
IS is supported by the Schweizerische Nationalfonds.
The calculations were made possible by grants
of computer time at the National Energy Research
Supercomputer Center.

\bibliographystyle{unsrt}
\bibliography{[sick.rev]sum2,[sick.deut]x}
\end{document}